\documentclass[twoside,twocolumn,english,aps,amsfonts,amsmath,prd,tightenlines, nofootinbib]{revtex4}
\usepackage[T1]{fontenc}
\usepackage[latin1]{inputenc}
\usepackage{amsmath}

\makeatletter

\usepackage{epsf}

\def\be{\begin{equation}}
\def\ee{\end{equation}}
\def\beq{\begin{equation}}
\def\eeq{\end{equation}}
\def\bea{\begin{eqnarray}}
\def\eea{\end{eqnarray}}
\def\bml{\begin{subequations}}
\def\blea{\bml\begin{eqnarray}}
\def\elea{\end{eqnarray}\end{subequations}}

\usepackage{babel}
\makeatother
\begin{document}

\title{Cosmic string loops: large and small, but not tiny}

\author{Vitaly Vanchurin}

\email{vitaly@cosmos.phy.tufts.edu}

\affiliation{Arnold-Sommerfeld-Center for Theoretical Physics, Department für
Physik, Ludwig-Maximilians-Universität München, Theresienstr.\ 37,
D-80333, Munich, Germany}

\begin{abstract}
We develop an analytical model to study the production spectrum of
loops in the cosmic string network. In the scaling regime, we find
two different scales corresponding to large (one order below horizon)
and small (few orders below horizon) loops. The very small (tiny)
loops at the gravitational back reaction scale are absent, and thus,
our model has no ultra-violet divergences. We calculate the spectrum
of loops and derive analytical expressions for the positions and magnitudes
of the small and large scale peaks. The small loops are produced by
large bursts of similar loops moving with very high velocities in
the same direction. We describe the shape of large loops, which would
usually consist of few kinks and few cusps per oscillation cycle.
We also argue that the typical size of large loops is set by the correlation
length, which does not depend on the intercommutation probability
$p$, while the interstring distance scales as $p^{1/3}$. 
\end{abstract}

\pacs{98.80.Cq 11.27.+d }

\maketitle

\section{Introduction}

Many cosmological models lead to the production of cosmic strings
as a result of the phase transition \cite{Kibble}, or cosmic superstrings
at the end of brane inflation \cite{Tye,Dvali,CMP}. Very early analytical
\cite{Kibble,Kibble:1984hp,Bennett:1986zn} and numerical \cite{AT,BB,AS}
investigations agreed that the evolution of infinite strings exhibits
scaling. An unexpected result came from the analysis of string loops
by means of numerical simulations, when the two groups \cite{BB,AS}
independently found that the loops do not scale, and are mostly produced
at the smallest resolution scale of the simulation. 

To tackle this problem an exact simulation for the cosmic string network
in flat space-time \cite{VOV} was developed. The precision of the
simulation was only set by unavoidable computer arithmetic uncertainty,
which is many orders of magnitude smaller than the scale of any structure
in the network. Using this code, it was shown that the production
of small loops is always accompanied by the production of somewhat
larger loops \cite{VOV2}. Later, it was conjectured \cite{OV} that
the production of small loops is a transient phenomena and in the
long run only the large loops are produced at the scales of the interstring
distance $\sim0.1t$. 

The two other groups \cite{RSB,MS} have recently performed the numerical
analysis with higher resolution versions of the simulation codes previously
used \cite{BB,AS} and were able to resolve the small loops. Reference
\cite{MS} found some evidence for a scaling distribution of loops,
but their size is apparently much smaller than in Refs. \cite{VOV2,OV}
and Reference\ \cite{RSB} found a distribution which diverges at
small sizes. We believe that the dynamical range of all three simulations
\cite{OV,AS,RSB} is not sufficient, to study the true scaling behavior.

In an attempt to settle the issue, in this article we develop an analytical
model of cosmic strings and compare it with another recently proposed
model of Ref. \cite{DPR,PR}. We show that in the final scaling regime
both types of loops (small and large) are present, and we do not see
any evidence of the tiny loops at the gravitational back reaction
scale.

The paper is organized as follows. In the next section we analyze
the production of large and medium-sized loops. In the third section
we study the production of small loops in the vicinity of cusps. We
show that the dominant scale of small loops is only few orders smaller
than horizon, and the production of loops at the gravitational back
reaction scale is strongly suppressed. In the forth section we calculate
the spectrum of loops in three different regimes with intercommutation
probability also taken into account. In the fifth section we derive
analytical expressions for the positions and magnitudes of the small
and large scale peaks in the power law approximation of the correlation
functions. In the last section we briefly discuss the effects of cross-correlations
between opposite-moving waves and compare our results with those obtained
in Ref. \cite{DPR,PR}.

\section{Large scales}

The evolution of cosmic strings can be captured by right ($\mathbf{a}(\sigma)$)
and left ($\mathbf{b}(\sigma)$) moving waves \[
\mathbf{x}(\sigma,t)=\frac{1}{2}(\mathbf{a}(\sigma-t)+\mathbf{b}(\sigma+t))\]
where $\mathbf{x}(\sigma,t)$ describes the shape of a given string.
When a self-intersection takes place, the following condition must
be satisfied\[
(\mathbf{\mathbf{a}}(\sigma+\Delta-t)\mathbf{-a}(\sigma-t))+(\mathbf{b}(\sigma+\Delta+t)-\mathbf{b}(\sigma+t))=0\]
where $\Delta$ is the size of a loop that could be formed, and $t$
is the time of the intersection. 

For convenience, we define \[
\mathbf{A}(\sigma)\equiv\int_{\sigma}^{\sigma+\Delta}\mathbf{a}'(x)dx=\mathbf{\mathbf{a}}(\sigma+\Delta)\mathbf{-a}(\sigma)\]
and \[
\mathbf{B}(\sigma)\equiv\int_{\sigma}^{\sigma+\Delta}\mathbf{b}'(x)dx=\mathbf{b}(\sigma+\Delta)-\mathbf{b}(\sigma)\]
The condition of self-intersection can be replaced by an inequality\[
\mathbf{A}(\sigma-t)+\mathbf{B}(\sigma+t)<\epsilon,\]
which must be satisfied for some $\sigma$ and $t$ in order for a
loop of a size from $\Delta$ to $\Delta+\epsilon$ to have a chance
to be formed. In the limit of small $\epsilon$, the self-intersection
takes place with probability approaching to one. If the above inequality
holds for an arbitrary small $\epsilon$, then we say that a loop
is formed with probability $p$, where $p$ is determined from the
underlying theory. 

In general, the vectors $\mathbf{A}(\sigma)$ and $\mathbf{-B}(\sigma)$
lie within a volume $V$ of a sphere with radius $\Delta$. However,
for the extreme scales corresponding to very large (Brownian) or very
small $\Delta$, the vectors are confined to a somewhat smaller sphere
or to a spherical shell respectively. In this article we model the
production of loops as a random process of finding $\mathbf{A}(\sigma)$
and $\mathbf{-B}(\sigma)$ such that their difference is smaller than
$\epsilon$. The intuition tells us, that it must be easier to find
the right counter-partner if the volume $V$ is smaller. However,
this is not the whole story and other factors must be taken into account
in order to avoid divergences at small scales.

If the vectors $\mathbf{A}(\sigma)$ and $\mathbf{-B}(\sigma)$ span
the entire sphere of radius $\Delta$, then we call loops formed at
that scale, the medium-sized loops and\[
V_{medium}=\frac{4}{3}\pi\Delta^{3}.\]
Large loops, are the loops formed from the vectors $\mathbf{A}(\sigma)$
and $\mathbf{-B}(\sigma)$, which are confined to a sphere of radius
$\propto\xi\sqrt{\frac{\Delta}{\xi}}$, where $\xi$ is the correlation
length along the string and \[
V_{large}=\frac{4}{3}\pi\Delta^{\frac{3}{2}}\xi^{\frac{3}{2}}.\]
Some of the large loops do not inter-commute with infinite strings
throughout their evolution, when other quickly rejoin back to the
network. As we are only interested in the loops that are not affected
by the infinite strings, it is reasonable to expect a sharp cut-off
at \begin{equation}
\alpha_{l}\sim\pi\xi,\label{eq:large}\end{equation}
for the interstring distance of order $\xi$ and intercommutation
probability $p=1$.

The very large loops are not stable due to their Brownian shape, regardless
of $p$. After few cycles of oscillations the loops fragment into
smaller loops, which can only decay further by gravitational radiation.
Therefore, the large loops in the final spectrum (after fragmentation)
are likely to consist of relatively straight segments with at most
few kinks ($\sim4$) and cusps ($\sim2$ per oscillation cycle), which
is in agreement with numerical simulations of Refs. \cite{VOV2,OV}.

\section{Small scales}

For small loops, the volume occupied by vectors $\mathbf{A}(\sigma)$
and $\mathbf{-B}(\sigma)$ is approximated by a spherical shell of
radius $\Delta$ and thickness $\delta$ with \[
V_{small}=4\pi\lambda_{1}^{2}\delta,\]
where\begin{eqnarray*}
\lambda_{1} & \equiv & \langle\mathbf{A}\rangle_{rms}\\
 & = & \langle\int_{\sigma}^{\sigma+\Delta}\mathbf{a}'(x)dx\rangle_{rms}=\\
 & = & \langle\mathbf{a}(\sigma+\Delta)-\mathbf{a}(\sigma)\rangle_{rms}\end{eqnarray*}
and\[
\delta\equiv\sqrt{\langle\mathbf{A}^{2}\rangle-\langle\|\mathbf{A}\|\rangle^{2}}.\]
Thus, nearly all of the small loops are emitted with very high velocities
by combining nearly straight segments of $\mathbf{a}$- and $\mathbf{b}$-waves
pointing in opposite directions. In such regions the following condition
is likely to be satisfied satisfied\[
\mathbf{a}'(\sigma)=-\mathbf{b'}(\sigma),\]
which indicates the presence of a cusp. 

Without loss of generality, we consider $\mathbf{a}$- and $\mathbf{b}$-segments
of the string such that \begin{eqnarray*}
\mathbf{a}'(0) & = & -\mathbf{b}'(0)\\
\mathbf{a}(0) & = & 0\\
\mathbf{b}(0) & = & 0\end{eqnarray*}
and expand it in powers of $\sigma$\begin{eqnarray*}
\mathbf{a}(\sigma) & = & \sigma\,\mathbf{a'}(0)+\frac{\sigma^{2}}{2}\,\mathbf{a''}(0)+\frac{\sigma^{3}}{6}\,\mathbf{a'''}(0)+...\\
\mathbf{b}(\sigma) & = & \sigma\,\mathbf{b'}(0)+\frac{\sigma^{2}}{2}\,\mathbf{b''}(0)+\frac{\sigma^{3}}{6}\,\mathbf{b'''}(0)+...,\end{eqnarray*}
where \[
\mathbf{a'}(0)\cdot\mathbf{a''}(0)=\mathbf{b'}(0)\cdot\mathbf{b''}(0)=0\]
due to the normality conditions $\|\mathbf{a'}(\sigma)\|=\|\mathbf{b}'(\sigma)\|=1.$

At the second order in $\sigma$, a loop is formed when

\begin{eqnarray*}
\mathbf{a}(\sigma_{1}-t)+\mathbf{b}(\sigma_{1}+t) & = & \mathbf{a}(\sigma_{2}-t)+\mathbf{b}(\sigma_{2}+t)\end{eqnarray*}
or\begin{eqnarray*}
(\sigma_{1}+\sigma_{2}-2t)\,\mathbf{a''}(0) & = & (\sigma_{1}+\sigma_{2}+2t)\,\mathbf{b''}(0),\end{eqnarray*}
which is satisfied only if $\sigma_{1}=\sigma_{2}=t$. Thus, the small
loops could not be formed for $t\neq0$ at the scales at which the
third derivative of $\mathbf{a}(\sigma)$ or $\mathbf{b}(\sigma)$
is negligible. 

The relevant statistical quantities are given by\begin{eqnarray*}
\lambda_{2} & \equiv & \langle\mathbf{A}'(\sigma)\rangle_{rms}\\
 & = & \langle\int_{\sigma}^{\sigma+\Delta}\mathbf{a}''(x)dx\rangle_{rms}=\\
 & = & \langle\mathbf{a}'(\sigma+\Delta)-\mathbf{a}'(\sigma)\rangle_{rms}\end{eqnarray*}
and \begin{eqnarray*}
\lambda_{3} & \equiv & \langle\mathbf{A}''(\sigma)\rangle_{rms}\\
 & = & \langle\int_{\sigma}^{\sigma+\Delta}\mathbf{a}'''(x)dx\rangle_{rms}=\\
 & = & \langle\mathbf{a}''(\sigma+\Delta)-\mathbf{a}''(\sigma)\rangle_{rms}.\end{eqnarray*}

The production of loops of size $\Delta$ in the vicinity of a cusp
is allowed only if the following condition is satisfied \[
\lambda_{3}(\Delta)\frac{\Delta^{3}}{6}\sim\lambda_{2}(\Delta)\frac{\Delta^{2}}{2}\]
There is a divergence at small scales in the total number of loops
if \[
\lim_{\Delta\rightarrow0}\frac{\lambda_{3}\Delta}{3\lambda_{2}}>0,\]
but the total energy emitted by a single cusp remains always finite.
In the absence of divergences the smallest size of loops $\alpha_{s}$
could be approximately obtained from \begin{equation}
\frac{\lambda_{3}(\alpha_{s})\alpha_{s}}{3\lambda_{2}(\alpha_{s})}\approx0.1,\label{eq:small}\end{equation}
which correspond to the smallest scale at which the third term in
the expansion of $\mathbf{a}(\sigma)$ remains of the same order as
the second term.

\section{Spectrum of loops}

Without loss of generality, in the reminder of the paper we calculate
the spectrum of loops at a fixed time $t=1$, assuming that the full
scaling was already established. This is done to remove a trivial
dependence of the scaling production function of loops on time $f(\frac{\Delta}{t})\rightarrow f(\Delta)$. 

In between the small ($\alpha_{s}$) and large ($\alpha_{l}$) cut-off
scales there are three different regimes in the production spectrum
of loops, which correspond to the populations of small, medium and
large loops. The production function of loops could be estimated as
\begin{equation}
f(\Delta)\Delta^{2}\propto\frac{\lambda_{2}(\Delta)}{V(\Delta)}\Delta^{2},\label{eq:spectrum}\end{equation}
as it must be inversely proportional to the volume $V$, where the
vectors $\mathbf{A}(\sigma)$ and $-\mathbf{B}(\sigma)$ live, and
directly proportional to the rate $\lambda_{2}$ at which the volume
$V$ is explored by these vectors. The two factors of $\Delta$ are
due to the fact that we are interested in the energy emitted into
logarithmic bins of the loop length.

From Eq. \ref{eq:spectrum} we obtain\begin{eqnarray*}
f_{small}(\Delta)\Delta^{2} & \propto & \frac{\lambda_{2}(\Delta)}{V_{small}(\Delta)}\Delta^{2}=\frac{\lambda_{2}(\Delta)}{4\pi\delta(\Delta)}\\
f_{medium}(\Delta)\Delta^{2} & \propto & \frac{\lambda_{2}(\Delta)}{V_{medium}(\Delta)}\Delta^{2}=\frac{3\lambda_{2}(\Delta)}{4\pi\Delta}\\
f_{large}(\Delta)\Delta^{2} & \propto & \frac{\lambda_{2}(\Delta)}{V_{large}(\Delta)}\Delta^{2}=\frac{3\lambda_{2}(\Delta)}{4\pi\xi^{\frac{3}{2}}}\sqrt{\Delta}.\end{eqnarray*}

The numerical analysis of Ref. \cite{thesis}, showed little dependence
of the intercommutation probability $p$ on the correlation length
$\xi$, which is a hidden effect of the cross-correlations to be discussed
in the last section. Deep in the scaling regime, when the vectors
\textbf{$\mathbf{A'}(\sigma)$} are \textbf{}correlated over large
distances, the structures undergo a multiple intersections. This leads
to a very high effective probability of their intercommutation regardless
of $p$. On the other hand, the nearby strings inter-commute with
very low probabilities, which causes the interstring distance to vary
with $p$. Let us assume that \[
d\propto\xi\, p^{\kappa},\]
 where $\kappa$ is yet unknown parameter. 

In the scaling regime the total energy density in long strings scales
as \[
\rho\equiv\frac{1}{d^{2}}\propto\frac{1}{\xi^{2}p^{2\kappa}},\]
but the effective intercommutation probability remains exactly one.
Therefore, the probability of a single intercommutation must be proportional
to the density of long strings, which implies that the production
functions of loops is \begin{equation}
f(\Delta)\Delta^{2}\propto p^{-2\kappa}\frac{\lambda_{2}(\Delta)}{V(\Delta)}\Delta^{2},\label{eq:spectrum_p}\end{equation}
where $p$ can be arbitrary small. 

For very low $p$, the loops of much large sizes are allowed to form
without ever rejoining back to the infinite network. Thus, the large
cut-off scale, $\alpha_{l}$, can be sensitive to the intercommutation
probability. A very large Brownian loop of size $\Delta$ would sweep
an area with radius $r\sim\xi\sqrt{\frac{\Delta}{\xi}}$ with about
$\frac{r^{2}}{d^{2}}$ strings passing through the area. Each of these
strings would intersect the Brownian loop roughly $\frac{r}{d}$ times,
and thus,

\begin{equation}
\alpha_{l}\sim\xi\, p^{\frac{2}{3}(3\kappa-1)}.\label{eq:large_p}\end{equation}
It follows that the total energy emitted into large loops is given
by\[
E_{total}\propto p^{-2\kappa}\int_{0}^{\xi\, p^{\frac{2}{3}(3\kappa-1)}}\frac{\lambda_{2}(\Delta)}{V_{large}(\Delta)}\Delta^{2}d\ln(\Delta),\]
At the same time, if the decay is dominated by production of large
loops, then $E_{total}\propto p^{-2\kappa}$. Therefore, the above
integral must not depend on $p$, which implies that $\kappa=\frac{1}{3}$
and \[
d\propto\xi p^{\frac{1}{3}}.\]
Remarkably, this result is in a very good agreement with numerical
simulation of Ref. \cite{thesis}, where $\kappa\approx0.31$. 

The above analysis shows that the cut-off scale of large loops, $\alpha_{l}$,
does not depend on $p$, and is always given by Eq. \ref{eq:large}.
The cut-off scale of small loops, $\alpha_{s}$, is determined from
the statistical properties of strings which was shown not to depend
on $p$. Since the production functions of loops scale as $p^{-2\kappa}=p^{-2/3}$
and the cut-off scales are independent of $p$, the over all spectrum
of loops must also scale as $p^{-2/3}$.

\section{Exponential approximation}

In what follows, we assume that the two-point correlation function
$C(\sigma)$ is approximated by\[
C(\sigma)=e^{-a|\sigma|}.\]
The first two parameters which describe the statistics of segments
$\mathbf{A}(\sigma)$ and $\mathbf{-B}(\sigma)$ are calculated exactly\begin{align*}
\lambda_{1}(\Delta) & =\sqrt{\int_{0}^{\Delta}e^{-a|\sigma|}dx\, dy}\\
 & =\sqrt{\frac{2\Delta}{a}-\frac{2}{a^{2}}(e^{-a\Delta}-1)}\\
\lambda_{2}(\Delta) & =\sqrt{2-2C(\Delta)}=\sqrt{a\Delta}\\
\lambda_{3}(\Delta) & =\sqrt{2a^{2}(1-e^{-a\Delta})}\end{align*}
and the third parameter is approximated in the limit of small loops\begin{align*}
\delta(\Delta) & =\sqrt{\lambda_{1}^{2}-(\int_{0}^{\Delta}C(\sigma))^{2}}=\sqrt{\frac{2a\Delta^{3}}{3}}.\end{align*}
Without the cross-correlations between opposite-moving waves taking
into account, one can make an estimate of the production function
of loops in the three regimes\begin{eqnarray}
f_{small}(\Delta)\Delta^{2} & \propto & \frac{p^{-2\kappa}}{4\pi}\sqrt{\frac{3}{2}}\Delta^{-1}\nonumber \\
f_{medium}(\Delta)\Delta^{2} & \propto & \frac{3p^{-2\kappa}}{4\pi}\sqrt{a}\Delta^{-\frac{1}{2}}\label{eq:three_spectrum}\\
f_{large}(\Delta)\Delta^{2} & \propto & \frac{3p^{-2\kappa}}{4\pi}\sqrt{a}\xi^{-\frac{3}{2}}\Delta\nonumber \end{eqnarray}

The energy scale of small loops can be estimated from Eq. \ref{eq:small}:
\[
\alpha_{s}\approx\frac{0.15}{a}.\]
There are two more relevant scales $\alpha_{m_{1}}$ and $\alpha_{m_{2}}$,
which specify the range of the medium loops. From Eqs. \ref{eq:three_spectrum}
$\alpha_{m_{1}}=(6\, a)^{-1}$ and $\alpha_{m_{2}}=\xi$.

An important ratio is given by\[
\beta\equiv\frac{\int_{\alpha_{s}}^{\alpha_{m_{1}}}f_{small}(\Delta)\Delta^{2}d\ln(\Delta)}{E_{total}}\]
where \begin{eqnarray*}
E_{total} & = & \int_{\alpha_{s}}^{\alpha_{m_{1}}}f_{small}(\Delta)\Delta^{2}d\ln(\Delta)+\\
 &  & \int_{\alpha_{m_{1}}}^{\alpha_{m_{2}}}f_{medium}(\Delta)\Delta^{2}d\ln(\Delta)+\\
 &  & \int_{\alpha_{m_{2}}}^{\alpha_{l}}f_{large}(\Delta)\Delta^{2}d\ln(\Delta),\end{eqnarray*}
which roughly determines the fraction of energy emitted into small
loops. For $\xi=0.1$, $a=3$ we obtain \begin{eqnarray*}
\alpha_{l} & \sim & 0.3,\\
\alpha_{s} & \sim & 0.05\end{eqnarray*}
and\[
\beta\sim0.5.\]
Thus, about $50\%$ of the energy goes into small loops in contrast
to $90\%$ obtained in Ref. \cite{Polchinski}, where the power-law
approximation of the two point correlation was assumed. Our model
with the power-law ansatz also leads to immediate divergences at small
scales. In a real network $\lambda_{3}$ can in principle be divergent
when the kinks are present. However, our analysis assumes the Taylor
expansion only around a given cusp. The probability of a near-cusp
region to contain recent kinks arbitrary close to the cusp is infinitely
small. %
\footnote{The old kinks are completely smoothed out by gravitational radiation
after a fixed (may be very large) time, but the frequency of cusps
formation decreases with time. Therefore, in the long run, the old
kinks can be completely ignored for the analysis of small loops formation
in the vicinity of cusps.%
} We conclude that physically relevant $\lambda_{3}$ cannot be divergent,
and $C(\sigma)$ should not be modeled by a power-law.

\begin{figure}
\begin{center}
\leavevmode\epsfxsize=3.5in
\epsfbox{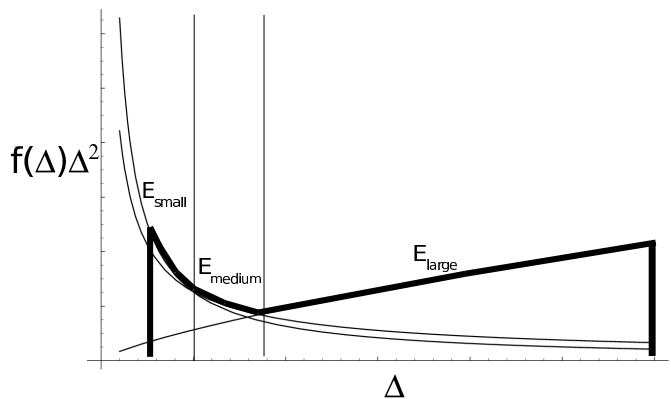}
\end{center}
\caption{Production function of loops.}
\label{fig:loops}
\end{figure}

The spectrum of loops for the above example is shown on Fig. \ref{fig:loops}. The qualitative shape of the plot agrees very well with numerical
simulation of Refs. \cite{VOV2,OV}. Nevertheless, it is too early
to expect a full quantitative agreement between our analytic model
and different simulations, due to a couple of reasons. First of all,
at present time all of the simulations are not entirely in the scaling
regime, as the production of loops is dominated by the remnants of
the initial conditions. Secondly, our model, as it stands, is ignorant
to the cross-correlations between opposite-moving waves. In the following
section, we will make a first attempt to include such effects into
our formalism.

\section{Discussion}

In the limit of negligible expansion the left and right moving waves
stay practically unchanged throughout the evolution. Therefore, \[
\mathbf{A}(0)-\mathbf{B}(\sigma)>\epsilon\]
must be satisfied for all $0<\sigma<\xi$, or otherwise a loop of
size $\Delta$ would have been formed already. Given that, we want
to find the probability $P$ that the following inequality holds:
\[
\mathbf{A}(0)-\mathbf{B}(0)<\epsilon.\]

At the moment, let us only consider the small loops. We expect $P$
to be smaller, than one would get without cross-correlations taken
into account, as it was done in the previous sections. On the other
hand, if another intercommutation is taking place such that \[
\mathbf{A}(\sigma)-\mathbf{B}(\sigma)<\epsilon\]
for some $-\xi<\sigma<0$, than $P$ must be much larger. The over
all effect on the production of small loops will be the same as if
no cross-correlations are present.

Although, the production of small loops is a very rare process as
the correlation length grows, the total effect is compensated by a
large number of loops produced at once. It follows that a number of
small loops are likely to be emitted in bursts, with very high velocities
pointing in approximately the same direction. These loops are not
likely to fragment further, since the third derivative of the segments
at even smaller scales ($<\alpha_{s}$) is not sufficiently large,
and as a result the small loops will decay only gravitationally. 

For the large and medium loops the situation is more complicated and
at present it is not clear how to take the effect of cross-correlations
into account. In addition, it is desired to include the intersections
of nearby strings, which inter-commute and produce kinks. All of these
effects complicate the problem dramatically and can in principle lead
to some unexpected results. At present we simply assume that the production
of large and medium-sized loops is captured entirely by the statistics
of long strings, in analogy to the small loops.

To conclude, we summarize the main differences of our results from
those obtained in Refs. \cite{DPR,PR}. In the scaling regime, we
do not expect to see very small loops at the scale of the gravitational
back reaction. Most of the energy is emitted into two different populations
of the string loops: small and large. The scale of large loops is
set by the correlation length, which does not depend on the intercommutation
probability. The scale of small loops is determined from the statistics
of structures on infinite strings, and is expected to be only few
orders lower than the scale of large loops. For illustration we used
an exponential form of the correlation function in contrast to the
power-law approximation used in Refs. \cite{DPR,PR}, which cannot
correctly represent a physically relevant statistics of strings at
small scales.

\section*{Acknowledgments}

We are grateful to Alex Vilenkin, Ken Olum and Joe Polchinski for
very helpful discussions and comments on the manuscript. This work
was supported in part by the Transregional Collaborative Research
Centre TRR 33 {}``The Dark Universe {}``.

\end{document}